\documentclass{iaus}
\usepackage{graphicx}

\title[Modeling the Disk (3-phase) ISM]
{Modeling the Disk (three-phase) Interstellar Medium}

\author[Gerhard Hensler]   
{Gerhard Hensler
}

\affiliation{Institute of Astronomy, University of Vienna,
Tuerkenschanzstr. 17, 1180 Vienna, Austria \\
email: hensler@astro.univie.ac.at}

\pubyear{2008}
\volume{254}  
\pagerange{119--126}
\date{?? and in revised form ??}
\jname{The Galaxy Disk in Cosmological Context}
\editors{J. Andersen, J. Bland-Hawthorn, \& B. Nordstroem, eds.}

\begin{document}
\def\HI{H{\sc i} }
\def\HII{H{\sc ii} }
\def\Ha{{\rm H}\alpha }
\def\SSF{\Sigma_{\rm SF} }
\def\Sg{\Sigma_{\rm g} }
\def\rSF{\rho_{\rm SF} }
\def\rg{\rho_{\rm g} }
\def\OIII{O[{\sc iii}] }
\def\Msun{M_{\odot} }
\def\Mpc2{\Msun/pc^2}
\def\tff{\tau_{\rm ff} }
\def\tSF{\tau_{\rm SF} } 
\def\e{\epsilon}
\def\eSF{\epsilon_{\rm SF} }
\def\OH{{12\-+log(O/H)} }
\def\lNO{log(N/O) }
\def\cd{chemo-dynamical }
\def \rmaa {Rev.Mex.\allowbreak  Astron.\allowbreak  Astrofis.}
\def\AA{A{\rm \&}A}

\maketitle

\begin{abstract}
The evolution of galactic disks from their early stages is dominated 
by gasdynamical effects such as gas infall, galactic fountains, and galactic 
outflows, and further more. The influence of these processes is only
understandable in the framework of diverse gas phases differing in 
their thermal energies, dynamics, and element abundances. 
To trace the temporal and chemical evolution of galactic disks, 
it is therefore essential to model the interstellar gasdynamics combined
with stellar dynamics, the interactions between gas phases, and 
star-gas mass and energy exchanges as detailed as possible. 
This article reviews the potential of state-of-the-art numerical schemes 
like Smooth-Particle and grid-based hydrodynamics as well as the
inherent processes as of star-formation criteria and feedback, energy 
deposit and metal enrichment by stars and on the influence 
of gas-phase interactions on the galactic gas dynamics and chemistry.
\end{abstract}

\firstsection 

\section{Introduction}

Galactic disks must have formed by dissipation and cooling of the galactic gas 
that remained after the first star-formation (SF) episode in the spheriodal 
component of galaxies and mixed with the returned and metal-enriched 
stellar matter. 
Cooling enforced the further collapse and angular-momentum kept the radially 
extended disk rotationally supported. Since then the disk must have been 
continuously fed and refreshed by infalling intergalactic gas but also depleted 
by galactic outflows.
As known from our Milky Way since the 70's and with the availability of 
formerly hidden spectral ranges (from the FIR/submm to X-rays) and also perceived 
in other disk galaxies the Interstellar Medium (ISM) consists of a multi-phase 
gas (\cite{fer01}), conditioning SF in the cool molecular components and 
driving the ISM dynamically mainly thru its very hot gas. 

Besides external effects structure and evolution of galaxies are primarily
determined by the energy budget of its ISM. Reasonably, also the ISM structure 
and its radiative and kinetic energy contents both are determined by the energy 
deposit of different sources, on the one hand, and by the energy loss by means 
of radiative cooling on the other.

The state of the ISM is thus not an isolated and meaningless questioning. 
Since stars are born from the cool gas phase but energize the ISM by means
of their energy and mass release, the treatment of the ISM must ultimately 
include also the stellar component with its distribution, age, and metallicity.
The disk formation and evolution thus depends intimately on the SF and
the energy content so that numerical simulations must account for these facts
properly. These are, however, the most challenging tasks for theoretical and 
numerical astrophysics of the ISM and are fundamental for galaxy evolution. 
But why?
On the one side, for SF rate in disks obeys the well-known Schmidt-Kennicutt 
law (\cite{ken98}) that relates the $\Ha$ surface brightness of disks with 
the gaseous surface density $\Sg$, and, on the other hand, 
we know from theory and empirical results how much energy is released by stars 
during their different evolutionary stages. So, where is the problem?
Is it not sufficient to relate the SF rate empirically to the gas mass
according to present galaxies in the local universe? 
Why should a detailed treatment of the ISM and its particular and temporal
composition be necessary and not a large-scale average of the ISM state?

In the following let us reflect and comment on the present state of the 
numerical treatment of the multi-phase ISM with its dynamics, its SF and 
energy feedback, its necessary small-scale physical interaction processes 
and the application to galaxy evolutionary models.
After having explained the most important effects of the ISM on galaxy
evolution and their empirical issues from observations, we will consider
representative numerical developments and question their drawbacks
and advantages without claiming for completeness. And at last we discuss 
the most advanced numerical efforts with respect to the requirements on a
most realistic approach. Nevertheless, already in advance let me 
concede that the influence of magnetic fields and of Cosmic Rays are 
not yet included and  not yet explored even to a first minor step  
in simulations.

\section{Structural and evolutionary effects on the Interstellar Medium}

\subsection{Star formation and the Schmidt-Kennnicutt Law}

Already in 1959 Schmidt argued that the SF rate per unit area relates to $\Sg$
by a power law with exponent $n$. The correlation found by Kennicutt (1998) of
\begin{equation}
\Sigma_{\Ha}[\Msun/(yr \cdot pc^2)] \propto \Sg [\Mpc2]^{1.4\pm 0.15}  
\end{equation}
\noindent
established that in equilibrium states the amount of gas determines the SF 
rate. Although this connection looked already pretty tight and holds over 
more than 4 orders of magnitude in $\Sg$ with a lower cut-off at almost 
10 $\Mpc2$, advancements by recent observations and theoretical interpretations
are more confusing. 
\cite{hey04} showed that this exponent is only valid for the molecular 
gas and that the correlation is tighter, while for pure \HI the slope 
is steeper. \cite{gao04} derived the dependence of the SF rate on the 
dense molecular cloud cores traced by the HCN molecule.

Such a Kennicutt relation can be theoretically simply supported as
\begin{equation}
\SSF = \frac{\Sg}{\tSF} \propto \frac{\Sg}{\rg^{-0.5}} \propto \Sg^{1.5}
\end{equation}
\noindent if the scaleheight H is independent on $\Sg$, i.e. the gas temperature 
is constant, so that $\Sg \propto \rho_{g,0} \propto \rg$. This fact can be 
only understood if heating and cooling both are independent of volume gas 
density $\rg$ or have the same dependence on it. 
K\"oppen, Theis, \& Hensler (1995) did however demonstrate already that the 
SF rate achieves a dependence on $\rg^2$, if the stellar heating is 
compensated by collisional-excited cooling radiation (according to \cite{bh89}).
The coefficient of this relation determines the SF timescale $\tSF$. 

But how can this be described? \cite{elm02} argued 
that the SF is determined by the free-fall time $\tff$. This, however, raises a 
conflict between the ISM conditions and observed SF rates in the sense that 
$\tff$ for a typical density of 100 cm$^{-3}$ amounts to 10$^{14}$ sec, 
i.e. $3\cdot 10^6$ yrs. For the galactic molecular cloud mass of 
$10^9 ... 10^{10}\, \Msun$ the SF rate should amount to 
$10^2 ... 10^3 \Msun$/yr what is by orders of magnitudes higher than observed! 
This means that the SF timescale
must be reduced with respect to collapse or dynamical timescale by introducing 
an efficiency  $\eSF$ and its definition could read: $\tSF = \eSF^{-1} \cdot \tff$.
Nevertheless, \cite{li05b} also obtained a Kennicutt law from numerical models 
of gravitationally unstable clouds slightly steeper than $n$=1.4.
 
Since this Schmidt-Kennicutt law is derived for gravitationally settled gas 
disks in rotational equilibrium, it should also reflect an equilibrium state 
of a self-gravitating vertically stratified ISM balanced by heating and cooling.  
It is therefore not surprising, that the exponent is larger for starburst 
galaxies, i.e. if the SF is e.g. triggered by dynamical processes, than in 
a quiescent mode so that the SF efficiency is increased.
This also explains why high-z galaxies reveal a steeper relation with an exponent $n = 1.7 \pm 0.05$ (\cite{bou07}), i.e. at stages when the gaseous 
disks form preferably by gas infall.  

Reasonably, one would expect that the Kennicutt relation can only be reproduced
in self-regulated disk models, i.e. in those where energy feedback by stars 
determines the energy budget of the ISM. \cite{kra03} identified in his 
cosmological simulations disk galaxies and analyzed them at different high 
redshifts. 
His model parameters spread in $\Sg$ over the large range observed and the 
model values scatter around the Kennicutt slope of $\SSF$ with $\Sg$. However, 
surprisingly, also those models without any energetic feedback by supernovae
(SNe) do not differ from those others with.
Similar conclusion can be drawn from the simulations by \cite{sdv08b} who 
neglect stellar feedback processes playing with parameters such as SF efficiency,
critical SF mass, etc. 

It is therefore still a matter of debate whether the Kennicutt-like dependence
of the surface SF rate is really a result of self-regulated SF in gas disk 
or if any dependences of self-regulation processes on the physical gas state 
cancel out so that the relation is a global one. 

For modelling the formation and evolution of galactic disks however the column
density is meaningless in yet dynamically developing systems. 
It is therefore necessary as long as the the disk has not achieved its 
equlibrium state of the gas and from the physical standpoint reasonable, 
to link the SF rate to the local state within a volume. 
(By the way, the same should hold for the modelling of gaseous dwarf galaxies 
(DGs) which are always in the state of re-arranging their gas structure 
because of the low gravitational potential.) This, however, means that 
the SF rate in a gas volume $\partial \rSF/ \partial t$ should be connected 
to the gas volume density $\rg$ with a power-law to the $m$. But is the
value of this $m$? Rana \& Wilkinson (1986) found a wide range of $m$ = 1 ... 4. 
Similarly to the above derivation of the Kennicutt exponent one can formulate 
\begin{equation}
\frac{\partial\rSF}{\partial t} = 
\frac{\rg}{\tSF} \propto \frac{\rg}{\rg^{-0.5}} \propto \rg^{1.5}
\end{equation} 

The proportionality again includes the SF efficiency. Such a 1.5 power-law
was already applied for the derived reason in the first gasdynamical simulations 
of galaxy evolution by Larson (1969 and in the 70's) and Burkert \& Hensler 
(1987, 1988). Under the assumption of triggered SF due to cloud-cloud collisions
a square dependence on $\rg$ should apply. In a simple picture, moreover, the cloud 
collision rate depends not only on the cloud number density but also on the
clouds' velocity dispersion and can by this directly affect the SF rate as a 
positive energetic feedback. Since dissipative collisions, however,
lead to a heating of the cloudy material, and if the SF rate depends on the
temperature of the cloudy ISM component, also a negative feedback is implied.
Therefore, a competition between collisional heating and gas cooling will determine
the sign of the SF feedback on energy and thus depends on the state and 
composition (metallicity, dust content, molecules, etc.) of the ISM.

Concerning the balance between direct stellar energy release to the
ISM and its collisionally-excited radiative cooling \cite{koe95} 
found that for the star-forming site itself self-regulation sets the dependence 
of the SF rate on $\rg^2$. In addition, the production of a SN-heated 
hot gas phase and its interaction with the cool gas by heat conduction leads
to a limit-cycle SF rate around a characteristic temperature due to the 
competition of evaporation and condensation (\cite{koe98}).

Consequently, the energy release and its efficiencies to transfer in into
dynamics and thermalisation are of evident importance. 
If their description and parametrization as well as a Schmidt-dependent
SF rate, respectively, are treated properly in numerical simulations, i.e.
close to real conditions, a Kennicutt dependence for a gaseous galactic disk
should be inherently reproduced.

\subsection{Stellar energy release and its transfer efficiency}

Because of their high power of released energies in various forms
massive stars are usually taken into account as the only heating
sources for the ISM. Moreover, a large fraction of simulations focusses
the energy deposit on SN typeII explosions alone. 
It must be discussed, however, how effective this energy is transferred
into the ISM as turbulent and consequently as thermal energy and an
even more sophisticated question must be expressed on the influence, 
transport and dissipation by Cosmic Rays and magnetic fields.

The same must be considered for the energy release of massive stars as
radiation-driven and wind-blown \HII regions.
Analytical estimates for purely radiative \HII regions yielded an energy
transfer efficiency $\e$ of the order of a few percent (\cite{las67}). 
Although the additional stellar wind power $L_w$ can be easily evaluated 
from models and observations, its fraction that is transferred 
into thermal energy ($\e_{th}$) or into turbulent energy is not obvious from first principles. Transfer efficiencies for both radiative and kinetic 
energies remain much lower than analytically derived from detailed numerical
simulations (more than one order of magnitude) and amount to only a few per mil 
(\cite{hen07} and references therein; Hensler et al. 2009, in preparation). 
Surprisingly, there is almost no dependence on the stellar mass (\cite{hen07}) 
as expected because the energy impact by Lyman continuum 
radiation and by wind luminosity both increase non-linearly with stellar mass. 
Vice versa, since the gas compression is stronger by a more energetic wind,
plausibly, also the energy loss by radiation is more efficient.

Another matter of fact is the energization of the ISM by SNe. Although it is 
generally assumed that the explosive energy lies around 10$^{51}$ ergs with 
an uncertainty (or intrinsic scatter) of probably one order of magnitude, 
SNeII do not explode as isolated and single events in the ISM. 
Almost without any exception massive stars do not disperse from the SF site
and thus explode within the stellar associations. 
Massive stars contribute significantly to this formation of structures 
such as e.g. cavities, holes, and chimneys in the \HI gas and superbubbles 
of hot gas (\cite{rec06a}).
On large scales the energy release by massive stars triggers the matter
circulations via galactic outflows from a gaseous disk and galactic winds.
By this, also the chemical evolution is affected thru the loss of 
metal-enriched gas from a galaxy (see e.g. \cite{rec06b}). 

SN explosions as an immediate consequence of SF stir up the ISM by the 
expansion of hot bubbles, deposit turbulent energy 
into the ISM and, thereby, regulate the SF again (\cite{hen02}). 
This negative energy feedback is enhanced at low gravitation because 
the SN energy exceeds easily the galactic binding energy and drives 
a galactic wind. This low gravitation also exists in vertical direction 
of a rotationally supported disk. 
Vice versa, SN and stellar wind-driven bubbles sweep up 
surrounding gas and can, by this, excite SF self-propagation as a 
positive feedback mechanism (e.g. \cite{ehl97,fuk00}). 

Since the numerical treatment of the galactic \cd evolution, even 
when it deals with two gas phases, cannot spatially resolve the ISM  
sufficiently, for a detailed consideration of small-scale processes
the \cd modeling has to apply parametrizations of 
plasmaphysical processes. It should, however, be repeatedly emphasized 
that an appropriate \cd description cannot vary those parameters arbitrarily, 
because they rely on results from theoretical, numerical, and/or 
empirical studies and are by this strongly constrained. 
This has been described and applied to different levels of \cd models 
(for an overview see \cite{hen03} and references therein, 
and more recently \cite{hen04a}; \cite{har06}).

Although investigations have been performed for $\e$ of SNe (\cite{tho98})
and superbubbles (e.g. \cite{str04}) they are yet too simplistic 
for quantitative results and the 10\% efficiency derived by
Thornton et al. is by far too large. 
Although numerical experiments of superbubbles and galactic winds in galaxies are performed,
yet they only demonstrate the destructive effect on the surrounding ISM but 
lack of self-consistency and complexity. 

Simulations of the chemical evolution of starburst DGs by 
\cite{rec06c} that are dedicated to reproduce the peculiar abundance patterns 
in these galaxies by different SF episodes found that $\e$ can vary widely. 
A superbubble which acts against the surrounding medium is cooling due to its
pressure work and radiation but compresses the swept-up shell material and
implies significant turbulent energy to the ISM. If a closely following SF
episode (might be another burst) pushes its explosive SNeII into the already
formed chimney, the hot gas can easily escape without a hindrance.
\cite{rec06a} found that depending on the external \HI density the chimneys do 
not close before a few hundred Myrs.

\subsection{Gas Infall}

Formation and evolution of galactic disks cannot be considered in isolation.
As confirmed from observations but also requested in the $\Lambda$CDM framework
galaxies are growing during their evolution by acquiring surrounding
gas continuously. Particularly, during the early formation phases it
remains, however, still unclear how this process works. Proposed are
cold and hot accretion (see e.g. \cite{ker05,wis07}), both dependent 
on mass and providing different inherent timescales of the galaxy growth,
but no clue exists yet which one dominates. 

In the local Universe several disk galaxies are enveloped by huge \HI halos 
and by refined velocity-position maps cold accretion in the form of \HI 
clouds with masses of about 10$^7 \Msun$ is manifested
(\cite{fra02}; Fraternali, this volume, and e.g. \cite{san08}). 
Since the infall rate determined by the detectable clouds fail to
support the SF rate in those galaxies by almost one order of magnitude,
it must be concluded that most of the infall happens by less massive clouds
perhaps like those observed in our Milky Way.

Although the element abundance properties of most Blue Compact DGs (BCDs) 
favour only a 
young stellar population of at most 1-2 Gyrs, they mostly consist of 
an underlying old population.
Furthermore, also most of these objects are embedded into \HI envelops 
from which at least NGC~1569 definitely suffers gas infall 
(\cite{sti02,mue05}). 
This has lead \cite{koe05} to exploit the influence of gas infall 
with metal-poor gas into an old galaxy with continuous SF on 
particular abundance patterns.
Their results could match not only the observational regime of BCDs in 
the [\OH-\lNO] space but also explain the shark-fin shape of the data
distribution. \cite{kna06} found also peculiar N/O abundance ratios in 
the galactic ISM that could also be explained by the infall scenario
of \cite{koe05}.

The accumulated explosion of massive stars lead to the formation of a 
superbubble which expands out of the gaseous disk, preferentially along the
steepest density gradient. In low-density environments when the thermal
energy does not exert too much PdV work or loose significant radiative 
energy, this expansion feeds the galactic halo with hot gas. On the other 
hand, \HI observations confirmed since a long time that high- und 
ultra-high velocity clouds of extragalactic origin fall towards the 
Galactic disk demonstrating that the halo consists of a multi-phase 
ISM (\cite{wol95}). 
Interstellar clouds with relative motion to surrounding gas are 
deformed and stretched by the ram pressure. \cite{hey96} exhibited
the cometary deformation of clouds exposed to the outflow of hot gas
in the SF region W4. Infalling clouds are not only influenced by a
drag force of the hot halo gas because of its relative velocity and 
by this decelerated (\cite{ben97}), also the thermal interaction between 
the gas phases by means of heat conduction affects their survival 
(\cite{vie07}). Due to their possible destruction and gas dispersion 
in hot environment, it is in general far from being clear how much of the 
infalling high-velocity clouds really reach the gaseous disk, feed the SF 
and excite turbulence, neither in the
early galaxy growth phase nor during the present evolution.

\subsection{Galactic Winds}

Some disk galaxies like e.g. NGC 253 and M82 not only supply hot halo gas
but, moreover, can drive a galactic wind. 
This scenario can be observed in various types of galaxies and in its 
different evolutionary stages. In the Galactic disk cavities and chimneys
become visible in \HI and in external galaxies \HI holes appear. 
In DGs superbubbles originate from super star clusters (SSCs) are observable by closed $\Ha$ loops and in X-rays still confined to the galaxy
like in NGC 1705 (see e.g. \cite {hen98}). NGC 1569 is a most interesting 
object: not only that gas infall occurs and probably triggers the present 
starburst (\cite{mue05}) but also a strong galactic wind exists. An analysis of 
the wind revealed a metal content around only solar to 2 times solar, 
i.e. much lower than expected from the yields by massive stars 
(\cite{mar02}). Since this low value requires a dilution with low-metallicity 
gas by a factor of about 10, a turbulent mixing alone in the shells 
of the expanding superbubble with the low-Z ISM in NGC 1569 seems insufficient. 
Most plausible is that gas clouds falling in from the envelopping \HI reservoir
 are evaporated and by this mass-loading the galactic wind.

\subsection{Structural issues}

In summary of the above mentioned observationally facts one has to be aware
that 

\begin{itemize}
\item the multi-phase structure of the ISM is co-existent in any galactic
regions,
\item the modes of SF and its positive or negative energetic feedback depend on
the small-scale physics of the ISM,
\item the reliability and significance of simple single-phase hydrodynamical 
models even with feedback are highly questionable.
\end{itemize}

\section{Numerical Methods}

\subsection{Star-formation prescriptions}

There are striking efforts to model the ISM on small scales in 
star-forming clouds in order to achive a deeper insight into the mechanisms
controlling SF (see e.g. \cite{ost01,mlkl04,li05,krk05}). Numerical
simulations on intermediate scales must already prescribe SF by some
reasonable recipes. 
Numerous papers (see e.g. \cite{wad99,avi04,sly05,tas08}) 
have implemented SF criteria such as e.g. 
threshold density $\rSF$ with SF if $\rg \ge \rSF$, 
excess mass in a specified volume with respect to the Jeans mass, 
i.e. M$_{\rm g} \ge M_{\rm J}$, 
convergence of gas flows ($div\cdot \underline{\bf v} < 0$), 
cooling timescale $t_{cool} \le t_{dyn}$, 
temperature limits T$_{\rm g} \le$ T$_{lim}$ , 
Toomre's Q parameter, 
and/or temperature dependent SF rate. 
If at least some of these conditions are fulfilled, as a further step
the gas mass which is converted into stars, i.e. the SF efficiency,
has to be set: fixing an empirical value $\eSF$ from observations, 
i.e. $\Delta m_{\rm SF} = \eSF \cdot \rg \cdot \Delta x^3$, where 
$\Delta x^3$ is the mesh volume in a grid code.
If the numerical timestep $\Delta t$ is smaller than the dynamical 
timestep $\tff$, $\Delta m_{\rm SF}$ has to be weighted by their time ratio 
(\cite{tas08}). 
Since $\eSF$ must inherently depend on the local conditions so that it is high 
in bursting SF modes, as requested for the Globular Cluster formation, but 
of percentage level in the self-regulated SF mode, numerical simulations
often try to derive the realistic SF efficiency by comparing models of largely
different $\eSF$ with observations (see e.g. \cite{tas08}: $\eSF$ = 0.05 and 0.5). 

While in many numerical treatments as the simplest prescription the Kennicutt 
law (\cite{sdv08}) is directly applied to the SF rate, another approach 
is the trial based in eq.(2.3) like 
\begin{equation}
\frac{\partial\rSF}{\partial t} = \alpha G^{1/2} \rg^{3/2} 
\end{equation}
\noindent and to express $\alpha$ using the Kennicutt law again and
applying an exponential disk with scaleheight $Z_0$ by
\begin{equation}
\alpha = 1.96 \cdot G^{-1/2} \frac{\SSF}{\Sg^{3/2}} Z_0^{1/2}
\end{equation}
\noindent But unreasonably, this is even applied to interacting galaxies 
(\cite{spr00}) where the application of both the Kennicutt law as well 
as the usual definition of the scaleheight, respectively, is highly 
questionable and to my opinion fails.

Theoretical studies by \cite{elm97} achieved a dependence of $\eSF$ on the
external pressure, while \cite{koe95} explored a temperature dependence of 
the SF rate.

\subsection{Smooth particle hydrodynamics}

Because of its simple numerical treatment and its inherent 3D representation 
{\it Smooth Particle Hydrodynamics} (SPH) is at present the most widely applied 
numerical strategy for simulations of cosmological structure formation, galaxy
formation, galaxy collisions and mergers, and further more. Because of its
particle character SPH simulations allow easily the simultaneous treatment of
the gas and the stellar component. The SPH subdivides the gaseous component in gas packages of sizes that represent the continuum character of diffuse gas, 
i.e. by large spherical extents of the SPH particles formulated by 
a kernel function $W_{ij}$ 
so that the mass density at particle i is given thru their neighbouring particles j with mass $m_j$ by 
$$
{\overline \rho_i}~=~\sum_{j=1}^N m_j~\cdot~W_{ij}.
$$

Although SPH is a quite powerful formalism for 3D problems, is has several 
drawbacks: 1) On the one hand, representing gas phases which change their 
scalelengths by orders of magnitude within a short range, i.e. adjacent
hot and cool particles, is impossible. Unfortunately, there are many 
authors who are misled by its applicability.
2) Ad hoc assumptions on SF and feedback are necessary. As an example: over which range, i.e. over how many adjacent particles has the SNII energy feedback 
to be taken into account? If the mass and spatial resolution of SPH particles 
is insufficient, one cannot avoid that the mean thermal energy of a gas 
particle does not increase due to SNeII, but remains at about 10$^4$ K
due to the efficient cooling. This coarse treatment is therefore insensitive 
to self-regulation, and, by this not accounting for a hot gas phase.
3) Multi-fluid descriptions are almost impossible. If one decouples
two kinds of particles totally, they can hardly experience the neighbourhood
of each other. Or one allows for different states but one kind of particle
so that each particle acts to its adjacent neighbour as a continuous fluid.

Since these weaknesses, which are mostly caused by the single gas-phase
description and the lack of interaction processes, 
have been recognized by many modellers, mainly also
in the field of cosmologigal simulations, numerous attempts to overcome
them have been performed, developing various strategies for the treatment of
a multi-phase ISM. Since {\sc Gadget} is probably the most widely distributed
and applied publicly available SPH code (\cite{spr00,spr01}) let us
consider preferably its application to disk galaxies and its advancements. 

\cite{dvs08} applied {\sc Gadget} to study the impact of SN feedback 
on galactic disks in order to drive a galactic wind and vice versa 
the effect of the wind strength on the SF. This latter is descriped by the 
Kennicutt law combined with a density threshold. For massive galaxies
with DM mass of 10$^{12} \Msun$ the mass resolution of gas particles 
reaches down to $5\cdot 10^4 \Msun$.
The SNII energy is deposited as kinetic wind energy and wind particles
are stochastically selected among neighbouring particles. In some models 
these are hydrodynamically coupled with the normal gas particles. 
As expected, the models showed that the galactic winds are stronger for 
larger energy release and, on the other hand, reduce the SF rate, but in
a not self-consistent manner because neither cold-gas infall nor SF 
trigger are included. In addition, in models with wind the SF rates 
decrease by about a factor of two already within 0.5 Gyrs, what is much 
too short for derived SF timescales of disk galaxies. 

Another recent development by \cite{sca06} uses a very sophisticated 
differentiation between the SPH particles into cold and hot ones but treats 
them still adequately except that particles with different thermodynamic 
variables (cold vs. hot) do not feel each other.
The recipe to allow for SNII energy to be distributed into the cold and hot
particles but not cooled away in the cold ones is the possibility of a 
so-called {\it promotion} of cold particles according to the state of their 
hot neighbors. \cite{sca06} claim that even mass-loaded galactic winds can 
develop in these models. Nevertheless, SF criteria and efficiency as well
as the distribution of the SN energy are freely parametrized and adapted
to galaxies in the local Universe observed in equlibrium states.

In order to describe the different ISM phases at least by two also
differently treated particles, \cite{sem02} introduced four different
particle species: warm gas as SPH particles, stars and DM as collisionless
particles, and the cold gas by means of sticky-particles. Mass exchange
between the cold and warm gas is limited by the transformation by means of
heating due to SN feedback vs. cooling. The cool particles are not intended 
to reprtesent a diffuse gas distribution but instead the clumpiness of
the cloudy ISM component and thus subject to individual cloud-cloud
collisions with only partial inelasticity. 

Further mass transfer between the components is allowed by SF and stellar 
mass loss. While the SF rate is described by a Schmidt law of the cool gas 
volume density with a power of 1 to 1.5, the mass fraction of formed stars  
is at first trapped within the cool gas particle. Not before a sufficient 
amount of stars exists in a specified number of neighbouring particles 
these stellar masses are subtracted from the gas particle and collected 
to a new pure stellar particle. \cite{sem02} applied an advanced procedure 
to keep the particle number almost constant. 
Further mass exchange is achieved by the stellar mass return according to
the stellar aging and death rate. The particles are numerically treated by
the so-called TREE algorithm (\cite{bar86}).

Although this approach looks well justified and explored, the figure
presentation of the best model over 2 Gyrs in \cite{sem02} shows that it
obviously suffers from an unrealisticly high SF rate. The gas is consumed
to 20\% within this short time and the SF rate has dropped from almost
100 $\Msun$/yr to less than 5. The corresponding e-folding timescale of 
SF amounts to only less than 0.5 Gyrs what stands in contradiction to 
the observationally derived ones e.g. in the Milky Way. 

This model behaviour of a too intensive SF might have different reasons:
no dynamical, energetical and materialistic interactions are implied into
the numerical treatment. Dynamically, the drag excerted by hot gas flow 
against cool clouds is not sufficiently represented by only the 
thermal pressure gradient built up by the local SN energy release and
driving the expansion of hot gas. From the energetical point of view,
heat conduction transfers energy from the hot to the cool clouds and,
by this, regulates the SF (\cite{koe98}). In addtion, this heat 
conduction also can lead to partial evaporation of clouds and thus reduce
the clouds mass reservoir available for SF.

Surprisingly, even recently the formation of exponential disks has 
been modelled by \cite{bou07} applying another speciality of algorithms,
a particle-mesh + sticky-particle scheme. Although they dealt with the
enormous number of one million particles for DM, gas and stars, their
limitation to a single gas phase without any stellar feedback are much
too severe to allow for reliable results, in particular, with respect to 
the aimed issues of this exploitation. The author aimed at understanding  
the evolution of clump-clusters and chain galaxies by fragmentation of
gaseous disks in gravitational unstable clumps. The reasoning for this
neglect that the feedback from SF ''is not well known'' and that the 
clump masses lie much above the influential range of SN energy is 
missleading as far as it has not been tested and proven that the 
coupling of energetic processes between scales is really inefficient. 

All the above mentioned additional processes, acting among the gas phases 
and, by this, coupling them even on different scales are additionally 
incorporated in the SPH/sticky particle treatment by \cite{har06}. 
Moreover, in their model the cloudy particles show an inherent mass 
distribution according to their stickyness and SF can happen in a 
''normal'' mode, but also induced by cloud collisions. In addition, to
the temperature criterion by \cite{koe95} the efficiency by \cite{elm97} 
steers the SF. As a result \cite{har06} demonstrated a stabilized 
rotating disk with moderate SF over 2 Gyrs, starting with a present-day
realistic value of 3 $\Msun$/yr for a gaseous disk and dropping off only 
by a factor of 4, what means a SF timescale of more than 1 Gyr. As argued 
by them, this decline in the SF rate is caused by a lack of external
gas that would be allowed to fall in as it happens to the Milky Way and
other disk galaxies and to keep on the SF. As a proxy, also the validity 
of the Kennicutt law could be almost represented but with a slope of -1.5.

The model treatment allows the mass release by massive and intermediate-mass 
stars, the latter as Planetary Nebulae (PNe), while the energy feedback is 
only attributed to SNeII by 20\% of the normally used SNII energy of
10$^{51}$ ergs. 
While until now both last mentioned numerical treatments do not include 
type Ia SNe, \cite{sem02} also trace the metal enrichment according to 
the stellar ejecta as a simple generalized Z, but \cite{har06} do not yet
although their prescription would allow for the differential treatment 
of abundance yields.

This task has been performed until now only in a further advanced development
by \cite{ber03,ber09} that is based on a single gas-phase SPH description 
by \cite{ber99}. The treatment is almost the same as in \cite{har06}, 
except that here the cool gas phase is also treated by SPH particles. 
This strategy can, however, be justified by the still large masses of the
gas packages of 10$^5\, \Msun$ at minimum i.e. in the range of 
Giant Molecular Clouds (GMCs) instead of small SF clumps. 
Therefore, also the cool gas can be considered as continuous. 
The gas phases can behave dynamically independently and are only 
coupled by drag and mass loading by means of heat conduction. 
For the SF the Jeans mass criterion is used again 

The most important advancement is the \cd treatment which includes elements 
released by different progenitor stars as PNe, SNeIa, SNeII, and stellar winds 
and what enables one to trace the galactic evolutionary epochs.

\subsection{Grid codes}

The advantages of particle code due to the easy numerical handling and
the simple inspectation is bought by the drawbacks of complication with
the multi-phase treatment of the ISM, the poor spatial resolution and 
with necessary physical processes acting and featuring the gas on 
various length scales. For these purposes, numerical codes based on 
spatial grids are at the moment more adaptable to resolution and more 
appropriate for the physics. Nevertheless, their full 3D treatment suffers
severely from spatial resolution. Because of the numerical costs, moreover
until now only a few codes exist which treat at least two separate gas
phases. Another serious problem is the treatment of stars. Two possibilities
exist: to incorporate the stellar component into the general fluid description
or to handle a hybrid code that switches between the hydrodynamics on the
grid to stellar dynamics.

Although many papers denote their published numerical applications as 
multi-phase gasdynamics (e.g. \cite{wad99, wad01}), the simulations 
in reality only deal with a single gas fluid and account for a range of 
gas temperatures and densities. \cite{sly05} studied the various effects
of SF, stellar feedback, self-gravity, and spatial resolution of the evolution
of the ISM structure in gaseous disks. The main conclusions which can be drawn
are that without SF the structure remains more diffuse than with and that
with SF stellar feedback is crucial, although at high spatial resolution the
SF within grid cells inherently looses its coherence.
The same is still continued in the, at present, spatially best resolved 
(magneto-)hydrodynamical simulations of the ISM by \cite{avi04} (and more
recent papers).  

Nevertheless, also in such grid simulations criteria of SF within a grid cell  
such as e.g. excess of a density thresholds, excess of Jeans mass, convergence 
of flows, cooling timescale lower than dynamical one, fall below threshold 
temperature (see e.g. \cite{tas08}) have to be applied. Furthermore, at
present all these applications lack of the direct interactions of different
gas phases by heat conduction, dynamical drag, and dynamical instabilities 
thru forming interfaces, but the most refined ones allow for resolving the
turbulence cascade.


In addition, such mixing effects at interfaces between gas phases and due to
turbulence (like e.g. those in the combined SWB/\HII complexes)
contribute to the observations by the enhancement or, respectively, dilution 
of metal abundances. Hydrodynamical grid models have therefore been 
performed that implied the released elements from nucleosynthesis in stellar
progenitors of different masses and therefore also of stellar lifetimes, 
so-called yields. Since also the metal-dependence of astrophysical and 
plasmaphysical processes should be properly included in this so-called 
chemo-dynamics, the combined effects of gasdynamics and chemistry can be traced. 

If the gasdynamics, including SF and stellar feedback, is coupled with the 
chemical evolution single-gas phase simulations in 1D are, however, misleading.
The reasons are that, at first, hot metal-enriched gas is mixed with cooler
gas and, secondly, the limitations to one dimension does not allow for
an escape of hot gas due to 2D shape effects. As a simple excercise, one
can mix 10$^4$ K warm gas with the released SNII explosions according to a SF
efficiency of a few percent what means that less than 1\% of hot gas
is contributed to the specific energy and momentum of the warm gas phase.
This results in two effects: the gas temperature does not exceed 10$^5$ K
and thus cools down almost instantaneously to 10$^4$ K again, 
i.e. the gas remains in its warm state; the gasdynamics of the warm gas 
dominate until the hot gas contributes a significant fraction
to the total energy density and the gas is already diluted by gas consumption 
so that the cooling is extended while the gas pressure drives an expansion.
Models of that kind (\cite{pip08}) should be considered with caution unless
two independent gas phases are treated and/or this at least in 2D. 

Since energetic effects affect the evolution of low-mass galaxies more 
efficiently than massive galaxies, their application to DGs is more 
spectacular and has been perfomed with a single gas-phase representation 
of the ISM but in 2D (e.g. by \cite{rec06b}) in order to reproduce the 
peculiar element abundances in Blue Compact DGs (\cite{rec06d}). Since gas infall is not only affecting the chemistry but 
also SF (\cite{hen04b}) and dynamics of outflows we have extended the former 
models by infalling clouds (\cite{rec07}) which are set into the
numerical grid not self-consistently as density enhancement. In these 
models, two kinds of cloud contributions are considered, only initially
existing and continuously formed, respectively. The issues are the following: 
due to dynamical processes and thermal evaporation, the clouds survive only 
a few tens of Myr. The internal energy of cloudy models is typically reduced 
by 20 -- 40 \% compared to models with a smooth density distribution.  
The clouds delay the development of large-scale outflows, helping
therefore in retaining a larger amount of gas inside the galaxy.  
However, especially in models with continuous creation of infalling clouds, 
their ram pressure pierce the expanding supershells so that through these holes
freshly produced metals can more easily escape and vent into the galactic wind.
Moreover, assuming for the clouds a pristine chemical composition, their 
interaction with the superbubble dilute the hot gas, reducing its metal 
content. The resulting final metallicity is therefore, in general, smaller 
(by $\sim$ 0.2 -- 0.4 dex) than the one attained by smooth models.

Starting from simple 2D HD models with metallicity inclusion (\cite{bur87},
1988) we developed a two-gas phase \cd prescription which was at first 
applied as 1D version to the vertical formation of the Galactic disk
(\cite{bur92}). 
The metallicity was treated as a general Z without differentiation of
particular elements between their stellar progenitors but with various
production timescales.
The model could successfully produce vertical stellar density and metallicity 
distribution while also providing the formation timescales of thin and 
thick disk components. 

This \cd scheme was further extended by us to 2D and to specific elements 
characteristic for different stellar mass progenitors and production lifetimes. This \cd treatment includes metal-dependent stellar yields and winds, SNeIa,
SNeII, PNe, metal-dependent cooling functions and heat conduction. The gas
phases are described by the Eulerian hydrodynamical equations, 
co-existing in every grid cell in pressure equilibrium, 
whereby the cool component as representative
of interstellar clouds is described with anisotropic velocity components
derived from the Boltzmann moment equations with source and sink terms from
cloud collisions. The SF is parametrized as explored in \cite{koe95} to 
guarantee self-regulation.

The application of the \cd code to the evolution of disk galaxies 
(\cite{sam96}) could successfully show for the Milky Way (\cite{sam97}) 
how and on which timescale the components halo, bulge, and disk formed 
and how the abundances developed, could represent the temporal evolution 
of gas abundances, of the radial stellar abundance gradients, of the 
abundance ratios, and of star-to-gas ratios. 
This also showed how many metals are stored in the hot gas phase and 
mixed with the cool one by evaporation of clouds and condensation. 

Up to now as the optimal grid code one can consider the further development
of the \cd scheme (\cite{sam97}) to 3D and with the stellar dynamics
for the stars. In addition, a cosmologically growing DM halo is included
into the simulations by \cite{sam03}. These models contain all the crucial
processes of SF self-regulation by stellar feedback, multi-phase ISM,
and temporally resolved stellar components according to the \cd 
prescription (\cite{hen03}, 2007). They cannot only trace 
the formation and evolution of the disk galaxies' components but also of
characteristic chemical abundances and are until now the best
self-consistent evolutionary models of disk galaxies because the disk
formation is included into the global temporal galaxy evolution. 
Moreover, the formation and evolution of disk galaxies in their infancy
were studied by these self-consistent simulations with respect to bulge 
formation and disk fragmentation (\cite{imm04}).

\section{Discussion}

While SPH codes appear plausibly to be the most appropriate numerical
treatment of galaxy evolution including gaseous disks, it must be
considered with caution that for the multi-component representation
with interaction processes
the particle number is still at present limited to a few 10000, what
means that gas packages and stellar particles in massive galaxies
represent unrealisticly large masses. Only for DGs this
effort leads to acceptable resolutions.

I am here not advocating against particle codes and I am optimistic, that
advancements in computer facilities and numerical codes will soonly 
allow to overcome the spatial resolution problem and allow for these 
multi-component descriptions. Nevertheless, at the moment grid startegies
look much more promising because small-scale processes are easier and more
reliably to be coupled with large-scale dynamics. In addition, physical 
processes can be treated more appropriately. This also allows an outlook 
if the self-consistent global models are further developed to include magnetic
fields, radiative and Cosmic-ray transport.

A word of care should be expressed with respect to empirical relations.
That agreements with the Kennicutt relation exist of simulations, which 
are totally differing in the treatment of feedback processes, while 
the $\Ha$ surface brightness relation deviate from the Kennicutt slope 
of 1.4 with atomic \HI gas, molecular gas, and for different modes of SF, 
any representation of this relation by any modelling of galaxy evolution 
seems to be too weak to validate any numerical treatment. 
Like this also metallicity gradients and age-metallicity relations in
galaxy disks seem not to serve as proxies for the validation of numerical
codes. This means, on the other hand,
 that issues of such relations and the request to disentangle inherent
evolutionary processes are overestimated and that these relations are robust
and insensitive to the intrisic state of galaxies.

Furthermore, the presented models show that it seems illusory to try 
to correlate SF with gas densities if feedback processes are neglected.

We are still far from fully self-consistent models that allow for the
inclusion of necessary astrophysical and plasmaphysical processes on
all length and time scales requested. Growing computer facilities will
allow soon for 3D AMR simulations but not only refined spatial and temporal
resolution is the clue, also coherence on different scales as necessary
for SF, turbulence, magnetic fields, etc. are requesting intuition,
brain waves, and a large fraction of hard code developing and testing.

\begin{acknowledgments}
The author is gratefully acknowledging collaborations and discussions 
on this topic with Peter Berczik, Dieter Breitschwerdt, Stefan Harfst, 
Stefan Hirche, Joachim K\"oppen, Simone Recchi, Andreas Rieschick, 
Markus Samland, Rainer Spurzem, Christian Theis, Wolfgang Vieser, 
and Herv\'{e} Wozniak. I also thank Simone Recchi for
his careful reading of the manuscript and the organizers cordially 
for their invitation to this conference and their support. 
This work is supported by the key programme ''Computational Sciences''
of the University of Vienna under project no. FS538001.
\end{acknowledgments}

\end{document}